\documentstyle[aps,pra,epsfig,twocolumn]{revtex}

\def\be{\begin{equation}}
\def\ee{\end{equation}}
\def\bea{\begin{eqnarray}}
\def\eea{\end{eqnarray}}
\def\bma{\begin{mathletters}}
\def\ema{\end{mathletters}}
\def\C{\hbox{$\mit I$\kern-.7em$\mit C$}}

\newcommand{\bra}[1]{\mbox{$\langle #1 |$}}
\newcommand{\ket}[1]{\mbox{$| #1 \rangle$}}
\newcommand{\braket}[2]{\mbox{$\langle #1  | #2 \rangle$}}
\newcommand{\proj}[1]{\ket{#1}\!\bra{#1}}

\begin{document}
\draft

\title{Reversible combination of inequivalent kinds of multipartite entanglement.}

\author{G. Vidal, W. D\"ur and J. I. Cirac}

\address{Institut f\"ur Theoretische Physik, Universit\"at Innsbruck,
A-6020 Innsbruck, Austria}

\date{\today}

\maketitle

\begin{abstract}
We present a family of tri-partite entangled states that, in an asymptotical sense, can be reversibly converted into EPR states shared by only two of parties (say B and C), and tripartite GHZ states. Thus we show that bipartite and genuine tripartite entanglement can be reversibly combined in several copies of a single tripartite state. For such states the corresponding fractions of GHZ and of EPR states represent a complete quantification of their (asymptotical) entanglement resources. More generally, we show that the three different kinds of bipartite entanglement (AB, AC and BC EPR states) and tripartite GHZ entanglement can be reversibly combined in a single state of three parties. Finally, we generalize this result to any number of parties. 

\end{abstract}


\narrowtext

 Understanding the inequivalent ways in which the parts of a composite system can be entangled is one of the central open questions of quantum information theory. When the system consists only of two parts, A and B, and it has been prepared in a pure state $\ket{\psi}_{AB}$, then its entanglement properties are qualitatively equivalent to those of an EPR state \cite{Ei35},
\be
\frac{1}{\sqrt{2}}(\ket{00}+\ket{11}),
\label{EPR}
\ee
in the following sense \cite{bipart}. If two parties, Alice and Bob, share $N$ copies of the system in state $\ket{\psi}_{AB}$ and are allowed to perform local operations assisted with classical communication (LOCC), then they can transform, reversibly in the large $N$ limit, the state of their systems into $N E(\psi_{AB})$ copies of an EPR state (\ref{EPR}), where $E(\psi_{AB})$ is the entropy of entanglement of state $\ket{\psi}_{AB}$ ---namely the von Neumann entropy of the reduced density matrix describing either part $A$ or $B$---. Thus, reversibility of asymptotical conversions justifies that we regard all bipartite pure-state entanglement as equivalent and quantify it by means of $E(\psi_{AB})$.

It has been recently shown \cite{multipart} that a GHZ state \cite{Gr89} 
\be
\frac{1}{\sqrt{2}}(\ket{000}+\ket{111}),
\label{GHZ3}
\ee
of a tripartite system is inequivalent, even in this asymptotic sense, to EPR states distributed among the parties. This indicates that there is genuine tripartite entanglement. Similarly \cite{multipart} (see also \cite{multipart2}), for any number $n$ of parties sharing a system, the $n$-partite GHZ state
\be
\frac{1}{\sqrt{2}}(\ket{0^{\otimes n}}+\ket{1^{\otimes n}}),
\label{GHZn}
\ee
can not be reversibly converted by means of LOCC into any distribution of entangled states each one involving less than $n$ parties. Here we will refer to entanglement of the form (\ref{GHZn}) as {\em canonical}. Thus, in the general case of a $m$-partite system, one can find {\em at least} $2^m-m-1$ kinds of entanglement that are asymptotically inequivalent. They correspond, for each $n = 2, \ldots, m$, to all $m!/(n!(m-n)!)$ kinds of $n$-canonical entanglement, that is involving different subsets of $n$ parties \cite{explanation}.

 In this Letter we show that inequivalent kinds of multipartite entanglement can be reversibly combined into a pure state by means of LOCC. More specifically, we show that all kinds of $n$-canonical entanglement, $n=2, \ldots, m$, can be combined in a $m$-partite state, and then again reextracted, with asymptotically vanishing losses.

 For instance, we will prove that $N$ copies of some tripartite state
\be 
\ket{\psi} \equiv c_0 \ket{000} + c_1 \ket{1}\frac{1}{\sqrt{2}}(\ket{11} + \ket{22}),
\label{example}
\ee 
can be reversibly transformed, in the limit $N\rightarrow \infty$, into copies of an EPR state shared by Bob and Claire, and copies of a GHZ state, that is,
\be
\ket{\psi}^{\otimes N}  \approx \ket{\mbox{EPR}_{BC}}^{\otimes Nl_{BC}} \otimes \ket{\mbox{GHZ}}^{\otimes Nl_{ABC}}.
\ee
This means that the asymptotical entanglement properties of $\ket{\psi}$ can be completely characterized by simply specifying the values of $l_{BC}$ and $l_{ABC}$. Some other states $\ket{\psi'}$ of three parties will be reversibly converted into the three inequivalent kinds of bipartite EPR states and GHZ states, so that their entanglement can be characterized by the multicomponent measure $L_{\psi'} \equiv (l_{AB}, l_{AC}, l_{BC}; l_{ABC})$. For an arbitrary number $m$ of parties, a $(2^m-m-1)$--component measure will also quantify the entanglement properties of some other states $\ket{\psi''}$, by providing the amount of all inequivalent kinds of $n$-canonical entanglement ($n=2,..., m$) that can be reversibly extracted from them. 

While it is not yet clear how many asymptotically inequivalent kinds of entanglement exist, not even whether there is only a finite number of them for the simplest non-trivial case ---i.e. for tripartite systems---, our results arguably help in the ongoing effort \cite{multipart,multipart2,rest} to understand and classify multipartite quantum correlations, as they show for the first time that it is possible to quantify the entanglement of a given state by relating it to several inequivalent forms of entanglement.

We start by analyzing the asymptotic properties of the tripartite state $\ket{\psi} \in \C^2\otimes\C^3\otimes\C^3$ given by equation (\ref{example}), where $c_0, c_1 \geq 0$, $c_0^2 + c_1^2 = 1$. First we will show how the parties can extract, from $N$ copies of it and by means of LOCC, up to  $Nc_1^2+g_2(N)$ EPR pairs between Bob and Claire and $NS(c_0^2,c_1^2) + g_1(N)$ GHZ states, where $S(\{x_i\}) = -\sum_i x_i\log_2x_i$ and both $g_1(x)/x$ and $g_2(x)/x$ vanish as $x\rightarrow \infty$. Then we will prove that the same amounts of canonical entanglement ---up to corrections that vanish as $N\rightarrow \infty$--- suffice to create the state $\ket{\psi}^{\otimes N}$ (actually a state as faithful to $\ket{\psi}^{\otimes N}$ as wished if $N$ can be made arbitrarily large). Therefore we will have $l_{BC} = c_1^2$ and $l_{ABC} = S(c_0^2, c_1^2)$. Finally, we will then describe generalizations of this result to all possible kinds of canonical entanglement and to an arbitrary number of parties.

Let us consider, then, two copies of $\ket{\psi}$, which after some convenient relabeling of the local orthonormal basis in ${\cal H}^{N=2}=\C^4\otimes\C^9\otimes\C^9$ can be written as 
\bea
\ket{\psi}^{\otimes 2} = c_1^2~&&\ket{1^0}\frac{1}{\sqrt{4}}(~\ket{1_1^0 1_1^0}+\ket{2_1^0 2_1^0}+\ket{3_1^0 3_1^0}+\ket{4_1^0 4_1^0}~)\nonumber\\
+~c_0c_1~&&\ket{1^1}\frac{1}{\sqrt{2}}(~\ket{1_1^1 1_1^1}+\ket{2_1^1 2_1^1}~)\nonumber\\
+~c_0c_1~&&\ket{2^1}\frac{1}{\sqrt{2}}(~\ket{1_2^1 1_2^1}+\ket{2_2^1 2_2^1}~)\nonumber\\
+~c_0^2~&&\ket{1^2}\ket{1_1^2 1_1^2}.
\label{2copies}
\eea

By means of a local measurement the parties can project the state (\ref{2copies}) into one of the three subspaces characterized by a constant coefficient $c_0^kc_1^{2-k}$ ($k=0,1,2$). We point out the relevant fact that either Alice, Bob or Claire could perform such a measurement locally because $\ket{\psi}$ is a linear combination $c_0\ket{\phi_1}+c_1\ket{\phi_2}$ of two locally orthogonal states $\ket{\phi_i}$, i.e. 
\be
\mbox{Tr} [\rho_i^{\alpha}\rho_j^{\alpha}] = 0 ~~~\forall i\neq j, ~~\alpha = A, B, C,
\label{locally}
\ee
where $\rho_i^{\alpha}$ is the reduced density matrix of $\ket{\phi_i}$ for subsystem $\alpha$, and this implies that the three subspaces of (\ref{2copies}) are also locally orthogonal; in other words, the parties can manipulate locally each of these subspaces independently. If the result of the measurement corresponds to $k=0$, then Bob and Claire will be sharing a $2^2$-level maximally entangled state, that is $2$ EPR pairs; if the outcome corresponds to $k=1$ then the parties end up sharing an EPR$_{BC}$ state and a GHZ state, as can be straightforwardly checked by expanding $\ket{EPR_{BC}}\otimes\ket{GHZ}$ as a linear combination of product states; finally, an outcome related to the subspace $k=2$ leaves the parties with a product state $\ket{000}$. 

This structure of outcomes easily generalizes to the case of $N$ copies. Let us call block $(N,k)$, denoted by $\ket{B_{N,k}}$, the normalized projection of $\ket{\psi}^{\otimes N}$ into the subspace characterized by the coefficient $c_0^kc_1^{N-k}$, that is
\be
\ket{\psi}^{\otimes N} = \sum_{k=0}^{N} c_0^{k}c_1^{N-k} \sqrt{b_{N,k}} \ket{B_{N,k}},
\label{expansion}
\ee
$b_{N,k} \equiv N!/[k!(N-k)!]$. A direct computation shows that $\ket{B_{N,k}}$ is of the form,
\bea
\ket{B_{N,k}} = \frac{1}{\sqrt{rt} }&\{& \nonumber\\
\ket{1^{k}}(~ \ket{1^{k}_1 1^{k}_1} + \ket{2^{k}_1 2^{k}_1} &+& \ldots + \ket{r^{k}_1 r^{k}_1} ~) \nonumber\\
+\ket{2^{k}}(~ \ket{1^{k}_2 1^{k}_2} + \ket{2^{k}_2 2^{k}_2} &+& \ldots + \ket{r^{k}_2 r^{k}_2} ~) \nonumber\\
\vdots && \nonumber\\
+\ket{t^{k}}(~ \ket{1^{k}_{t} 1^{k}_t} + \ket{2^{k}_t 2^{k}_t} &+& \ldots + \ket{r^{k}_t r^{k}_t} ~)~~\},
\label{block}
\eea
where $r\equiv 2^{N-k}$, $t\equiv b_{N,k}$ and the local states satisfy $\braket{i^{k}_a}{j^{k'}_b} = \delta_{i,j}\delta_{k,k'}\delta_{a,b}$ in both Bob and Claire and $\braket{i^{k}}{j^{k'}} = \delta_{i,j}\delta_{k,k'}$ in Alice. Notice that (\ref{block}) is equivalent to the tensor product of a $r$-level EPR state and a $t$-level GHZ state,
\bea
\ket{B_{N,k}} = (\frac{1}{\sqrt{r}}\sum_{i=1}^r \ket{ii}_{BC})\otimes(\frac{1}{\sqrt{t}}\sum_{i=1}^t \ket{iii})\nonumber\\
= \ket{r\!-\!EPR_{BC}}\otimes\ket{t\!-\!GHZ}.
\label{block2}
\eea
Thus, by means of a local measurement projecting onto these blocks, the parties will obtain state (\ref{block2}) with probability $P_{N,k} \equiv  b_{N,k}c_0^{2k}c_1^{2(N-k)}$. The expectation values 
\bea
<\frac{\log_2 r}{N}> &=& \sum_{k=0}^N P_{N,k} \frac{N-k}{N}, \nonumber\\
<\frac{\log_2 t}{N}> &=& \sum_{k=0}^N P_{N,k}  \frac{\log_2 b_{N,k}}{N}
\label{expect} 
\eea  
correspond, respectively, to the fraction $l_{BC}$ of EPR$_{BC}$ states and to the fraction $l_{ABC}$ of GHZ states that are obtained, on average and per copy of $\ket{\psi}$, from $\ket{\psi}^{\otimes N}$ \cite{levelGHZ}. Because of the smooth behavior of the functions $(N-k)/N$ and $(\log_2 b_{N,k})/N$ compared to the binomial distribution $P_{N,k}$, we can calculate the expectation values (\ref{expect}) ---up to corrections that vanish in the limit $N \rightarrow \infty$--- by just evaluating the two functions at the peak of $P_{N,k}$, namely at $k_{\max} \equiv Nc_0^2$, which gives us the announced amount of entanglement for each of the two canonical forms.

Let us look now at the inverse transformation. We start, for clearness sake, by showing that $2$ EPR$_{BC}$ and $2$ GHZ states suffice to create state (\ref{2copies}) locally and with certainty. Let us expand $\ket{EPR_{BC}}^{\otimes 2}\otimes\ket{GHZ}^{\otimes 2}$ as
\bea
\frac{1}{\sqrt{4}}~\{~&&\ket{1^0}\frac{1}{\sqrt{4}}(~\ket{1_1^0 1_1^0}+\ket{2_1^0 2_1^0}+\ket{3_1^0 3_1^0}+\ket{4_1^0 4_1^0}~)\nonumber\\
+~&&\ket{1^1}\frac{1}{\sqrt{4}}(~\ket{1_1^1 1_1^1}+\ket{2_1^1 2_1^1}+\ket{3_1^1 3_1^1}+\ket{4_1^1 4_1^1}~)\nonumber\\
+~&&\ket{2^1}\frac{1}{\sqrt{4}}(~\ket{1_2^1 1_2^1}+\ket{2_2^1 2_2^1}+\ket{3_2^1 3_2^1}+\ket{4_2^1 4_2^1}~)\nonumber\\
+~&&\ket{1^2}\frac{1}{\sqrt{4}}(~\ket{1_1^2 1_1^2}+\ket{2_1^2 2_1^2}+\ket{3_1^2 3_1^2}+\ket{4_1^2 4_1^2}~)\}.
\label{2x2}
\eea
 In this expression (cf. (\ref{2copies})) we would like to give the first row a weight $c_1^2$; in both the second and third rows we should get rid of $\ket{3^2_i 3^2_i}$ and $\ket{4^2_i 4^2_i}$ and give each of the rows a weight $c_0c_1$; the fourth row should be reduced to a product state with weight $c_0^2$. After these changes we will have state (\ref{2copies}). We first note that the parties can transform, with certainty, the $2$ GHZ states into a triorthogonal state with arbitrary relative weights,
\be
\frac{1}{2} \sum_{i=1}^4 \ket{iii} \longrightarrow \sum_{i=1}^4 \lambda_i \ket{iii},
\label{transform}
\ee
by means of a local POVM $\{O_j\}$, $j=1,\ldots 4$, on (any) one of the parties followed by an outcome dependent, local unitary $U_j$ applied once in each of the parties' subsystems. Here $O_j$ is proportional to $\sum_i \lambda_i \proj{i\oplus_4 j}$ \cite{oplus} and $U_j$ takes $\ket{i\oplus_4 j}$ into $\ket{i}$ on each local subsystem. The tensor product of $2$ EPR$_{BC}$ states with the resulting state in (\ref{transform}) is equivalent to (\ref{2x2}) but with row $i$ having weight $\lambda_i$. Bob and Claire can now address each of the $4$ rows locally and reduce their length at wish. Indeed, in order to shorten the fourth row into a product vector, one of them, say Bob, can perform a POVM with $4$ positive operators 
\be
Q_i = \proj{i^2_1} + \frac{1}{2} \sum_{j=2,3,4} P_j, ~~~i=1,\ldots 4,
\label{POVM1}
\ee
where $P_j$ is a projector onto the local subspace supporting row $j$, e.g. $P_2 = \sum_i \proj{i_1^1}$; then Bob and Claire need to relabel the surviving term  $\ket{i^2_1i^2_1}$ of the first row as $\ket{1^2_11^2_1}$. By means of similar POVMs followed by outcome dependent local unitaries Bob and Claire can tailor also the second and third row so that they contain only $2$ product terms each. Explicitly, a $2$-outcome POVM that reduces the second row reads
\bea
Q'_1 = \proj{1^1_1  1^1_1}+\proj{ 2^1_1 2^1_1} + \frac{1}{\sqrt{2}} \sum_{j=1,3,4} P_j,\nonumber \\
Q'_2 = \proj{3^1_1  3^1_1}+\proj{ 4^1_1 4^1_1} + \frac{1}{\sqrt{2}} \sum_{j=1,3,4} P_j.
\label{POVM2}
\eea
 Notice that such measurements do not modify the relative weight of the rows. A proper choice of the coefficients $\lambda_i$ in the first step of the transformation, namely $\lambda_1=c_1^2, \lambda_2=\lambda_3=c_1c_0$ and $\lambda_4=c_0^2$, completes therefore the protocol for preparing $2$ copies of $\ket{\psi}$.

 In the case of a large number $N$ of copies the parties start with several EPR$_{BC}$ and GHZ states and want to create a state
\be
\ket{N_{k_-}^{k_+}} \equiv K\sum_{k=k_-}^{k_+} c_0^{N-k}c_1^{k} \sqrt{b_{N,k}} \ket{B_{N,k}}
\label{approx}
\ee
such that $F\equiv|\braket{N_{k_-}^{k_+}}{\psi^{\otimes N}}|^2= 1 - h(N)$, where $h(x\rightarrow \infty)=0$, that is a state which asymptotically can not be distinguished from $\ket{\psi}^{\otimes N}$. First we note that an arbitrary faithfulness can be achieved, asymptotically, by considering only the blocks $\ket{B_{N,k}}$ (cf. eq. (\ref{block})) that correspond to $k$'s around $k_{\max} = Nc_0^2$. Indeed, a straightforward computation of the fidelity shows that it suffices to take $k_\pm=c_0^2N \pm \alpha N^\beta$ for some $\alpha >0$ and $1/2 < \beta <1$: using that a binomial distribution is asymptotically equivalent to a normal (Gaussian) distribution, the fidelity $F$ can be seen to be bounded from below by $\Phi(2\alpha N^{\beta - 1/2})$, where $\Phi(x)\equiv1/\sqrt{2 \pi} \int_{-x}^x e^{y^2/2}dy$. For our choice of $\alpha, \beta$, we have that $F \rightarrow 1$ when $N \rightarrow \infty$. As with the $N=2$ case, our plan is, starting from a reasonable amount of EPR$_{BC}$ and GHZ states ---which can be expanded in the fashion of (\ref{2x2})---, ($i$) to modify conveniently the weight of each row in the expansion and ($ii$) to shorten the length of each row, to obtain the pattern of lengths given by the block structure of (\ref{approx}). Notice that a straightforward generalization of (\ref{transform}) provides row $i$ with an arbitrary weight $\lambda_i$ by locally manipulating the initial $GHZ$ states, and that we have also already seen how to shorten each row independently by means of a local POVM in either Bob's or Claire's side (see POVMs (\ref{POVM1}) and (\ref{POVM2}) as examples). Thus, the only question left concerns the amount of canonical entanglement required to produce all blocks $\ket{B_{N,k}}$ in (\ref{approx}). Since we have a mechanism to shorten but not to lengthen the rows, the number of EPR$_{BC}$ states must allow to obtain the longest rows, which are those of the block $\ket{N,k_-}$ and contain $2^{N-k_-}$ product terms each. That is, $N-k_-$ EPR$_{BC}$ states will suffice. The total number of GHZ states required is the logarithm of the total number of rows in (\ref{approx}), and thus reads $\log_2(\sum_{k=k_-}^{k_+}b_{N,k})$. Let $k_0 \in [k_-, k_+]$ be the value that maximizes $b_{N,k}$ in this interval. Then the amount of GHZ states required to prepared (\ref{approx}) is bounded from above by $\log_2 [(k_+-k_-+1)b_{N, k_o}]$. Finally, substitution of $k_{\pm}$ and $k_0$ in this bound and the previous estimation for EPR$_{BC}$ states shows that both amounts of canonical entanglement needed to prepare (\ref{approx}) are the expected ones, apart from corrections which scale sublinearly in $N$ and that therefore become irrelevant for $N$ sufficiently large. This concludes the proof that state (\ref{example}) is asymptotically equivalent to canonical entanglement.

We can now generalize the previous example and reversibly combine the three kinds of bipartite entanglement and the canonical tripartite entanglement in a single state. Indeed, the tripartite state
\bea
\ket{\psi'} \equiv &&c_0 \ket{000}\nonumber\\
+ &&c_1 \ket{1}\frac{1}{\sqrt{2}} (\ket{11}+\ket{22})\nonumber\\
+ &&c_2 \frac{1}{\sqrt{2}}(\ket{233} + \ket{334})\nonumber\\
+ &&c_3 \frac{1}{\sqrt{2}}(\ket{44}+\ket{55})\ket{5},
\eea
can be transformed by means of LOCC, into EPR states shared by $2$ parties and into GHZ states, the asymptotic ratios being, $l_{AB} = c_1^2$, $l_{AC} = c_2^2$ and $l_{BC} = c_3^2$ for the bipartite entanglement and $l_{ABC} = S(\{c_i^2\})$ for the tripartite entanglement. 

This result follows from considering analogous transformations to the ones described above. The expansion of the state of $N$ copies of $\ket{\psi'}$ into locally orthogonal subspaces as in (\ref{expansion}) depends now on $3$ independent indices, the weights defining the block structure being $c_0^{k_0}c_1^{k_1}c_2^{k_2}c_3^{N-k_1-k_2-k_3}$. The binomial probability distribution is replaced by a multinomial distribution ---centered at $k_i=Nc_i^2$--- and each one of the new blocks $\ket{B_{N,k_0,k_1,k_2}}$ is equivalent to the tensor product of some number of GHZ, EPR$_{AB}$, EPR$_{AC}$ and EPR$_{BC}$ states. A local measurement onto the blocks leads, for sufficiently large $N$, to the desired expectation values for the fractions of canonical entanglement. Conversely, these amounts of entanglement suffice to create all the relevant blocks $\ket{B_{N,k_0,k_1,k_2}}$. This is done by introducing some weights $\lambda_i$ in the initial GHZ states and by locally tailoring the (now multidimensional) rows in the pertinent expansion, as we explained in the previous example.

More generally, let the $m$-partite state 
\be
\ket{\psi''} \equiv \sum_{i=0}^l c_i \ket{\phi_i}
\label{general}
\ee
be a linear combination of locally orthogonal states [see equation (\ref{locally})] such that each one is the tensor product of a canonical state $\ket{\tau_i}$ (\ref{GHZn}) for $n_i$ of the parties, and a product vector for the remaining $m-n_i$ parties \cite{example}. Then $N$ copies of the state $\ket{\psi''}$ are asymptotically equivalent to $Nc_0^2$ copies of $\ket{\tau_0}$, $\ldots$, $Nc_l^2$ copies of $\ket{\tau_l}$ and to $NS(\{c_i^2\})$ copies of a $m$-canonical state, i.e.
\be
\ket{\psi''}^{\otimes N} \approx [\bigotimes_{i=0}^l \ket{\tau_i}^{\otimes Nc_i^2}] \otimes (\ket{0^{\otimes m}} + \ket{1^{\otimes m}})^{\otimes NS(\{c_i^2\})}.
\ee

 In this Letter we have provided examples of multipartite states whose entanglement properties can be classified and quantified in relation to the set of canonical states (\ref{GHZn}). The criteria underlying such classification is the asymptotical equivalence of states under LOCC. We have shown that at least in some cases, namely for states of the form (\ref{general}), this criteria brings a significant simplification in the general problem of classifying entanglement. Indeed, our results show that the states (\ref{general}), which depend on the set of continuous non-local parameters $\{c_i\}$, contain only a finite set of inequivalent forms of entanglement. We have gone further and quantified the amount of each form of entanglement contained in state (\ref{general}), which gives rise to a multicomponent measure. We do not know to what extend the coefficients, as well as the reference states of this measure are essentially unique.

This work was supported by the Austrian Science Foundation under the SFB 
``control and measurement of coherent quantum systems'' (Project 11), the 
European Community under the TMR network ERB--FMRX--CT96--0087, the European Science Foundation and the Institute for Quantum Information GmbH. G.V also acknowledges a Marie Curie Fellowship HPMF-CT-1999-00200 (European Community).



\begin{references}
\bibitem{explanation} 
 In \cite{multipart} it has been shown that $m$-canonical entanglement is asymptotically inequivalent to any distribution of different kinds of canonical entangled states if each of them involves less than $m$ parties. We can easily argue that also $n$-canonical entanglement ($n<m$) involving parties $\alpha_1, \ldots, \alpha_n$ is actually inequivalent to a distribution of any other kinds of canonical entanglement. Indeed, we need only notice that for the transformation to be possible, the distribution of final states can only contain entanglement between parties $\alpha_1, \ldots, \alpha_n$. But in this case the result of \cite{multipart} applies, because our original state is the $n$-canonical entangled state in a subsystem of $n$ parties and the rest of candidates in the final distribution must involve less than $n$ parties.
\bibitem{levelGHZ} A $r$-level $n$-canonical state is a $n$-partite state of the form $\frac{1}{\sqrt{r}}\sum_{i=1}^r \ket{i^{\otimes n}}$. For any integer $l \leq \log_2 r$ this state can always be transformed, locally and with certainty, into $l$ copies of a ($2$-level) $n$-canonical state (\ref{GHZn}) by means, for instance, of a local POVM on one of the parties consisting of $2^l$-dimensional positive operators $O_j$, $j=1,\ldots, r$ proportional to $\sum_{i=1}^{2^l} \proj{i\oplus_r j}$ \cite{oplus} followed by the outcome dependent local unitary $\ket{i\oplus_r j} \longrightarrow \ket{i}$ performed by each party.
\bibitem{example} For instance, for $m=5$, we could have $n_0=3$ and $\ket{\phi_0} = \ket{0}1/\sqrt{2}(\ket{000}+\ket{111})\ket{0} = \ket{0}\ket{\tau_0}\ket{0}$.
\end{references}
\end{document}